\documentclass{blois}

\def\be{\begin{equation}}
\def\ee{\end{equation}}
\def\bea{\begin{eqnarray}}
\def\eea{\end{eqnarray}}

\newcommand{\Photo}{}

\begin{document}
\vspace*{4cm}
\title{AUTOMATISED COMPUTATIONS OF ELECTROWEAK CORRECTIONS USING SHERPA+RECOLA}

\author{ MATHIEU PELLEN }

\address{ Universit\"at W\"urzburg, Institut f\"ur Theoretische Physik und Astrophysik, \\  
        97074 W\"urzburg, Germany}

\maketitle\abstracts{
Given the experimental precision and the multitude of processes that can be measured at the LHC, electroweak (EW) corrections become more and more relevant.
This calls for the automatisation of EW corrections and in particular their implementations in multi-purpose Monte Carlo generators in order to allow for a systematic study of such contributions.
In these proceedings I present the implementation of the matrix-element generator Recola in the Sherpa Monte Carlo.
The combination Sherpa+Recola allows to compute, in principle, any process in the Standard Model at NLO QCD and EW accuracy.
In addition to fixed-order NLO computations, all functionalities of Sherpa such as parton shower or hadronisation are still available.
}

\section{Introduction}

With the increasing experimental precision at the LHC, precise and appropriate predictions are more than ever needed.
In addition to QCD corrections that should be computed at least at next-to-leading order (NLO) in combination with parton shower, electroweak (EW) corrections should also be taken into account.
Indeed, with the increase in energy and the accumulated data, experimental collaborations will be able to probe the tails of distributions of many processes.
In this very region, EW corrections are particularly relevant due to the appearance of Sudakov logarithms that become negatively large.

In order to compute NLO EW corrections, the first ingredient needed is a matrix-element generator able to provide EW one-loop amplitudes.
Several of these programs are now available~\cite{Cullen:2014yla,Cascioli:2011va,Hirschi:2011pa,Actis:2016mpe}.
The second ingredient is a Monte Carlo program that combines the virtual and real corrections.
In addition, several Monte Carlos provide theoretical descriptions beyond the hard matrix element such as parton shower or hadronisation.
This allows a realistic description of event collisions and the systematic study of NLO EW corrections.
Thus, the implementation of general matrix-element generators in multi-purpose Monte Carlo code is essential as it allows theorists and experimentalists to compute arbitrary processes at NLO QCD and EW accuracy.

One recent progress in the automation of EW corrections is the implementation of the matrix-element generator 
Recola\footnote{For the numerical evaluation of the one-loop scalar and tensor integrals, Recola relies on the COLLIER library~\cite{Denner:2014gla,Denner:2016kdg}. }~\cite{Actis:2016mpe,Actis:2012qn}
in the Sherpa Monte Carlo~\cite{Gleisberg:2008ta,Gleisberg:2003xi} which is dubbed Sherpa+Recola~\cite{Biedermann:2017yoi}.
As the implementation is completely general, it allows to calculate any process in the Standard Model with NLO QCD and EW accuracy.
In the following, I present examples of NLO EW computations made in this new implementation.
This ranges from the on-shell production of a Higgs boson in association with on-shell top quarks, to the off-shell production of two Z bosons, to the production of a gauge boson in association with several jets.
These processes all have different characteristics with respect to NLO computations, demonstrating hence the generality of Sherpa+Recola.

For explanations concerning the implementation, technical details, and more results, the interested reader is referred to the original article presented in Ref.~\cite{Biedermann:2017yoi}.
Note that both Sherpa\footnote{Sherpa is publicly available at \url{https://sherpa.hepforge.org}.} and Recola\footnote{Recola is publicly available at \url{https://recola.hepforge.org}.} can be downloaded and readily be used to compute NLO QCD corrections.
The NLO EW functionalities of the interface will soon be made public.

\section{Example processes}

\subsection{${\rm p} {\rm p} \to {\rm t} \bar {\rm t} {\rm H}$}

\begin{figure}
 \begin{center}
  \includegraphics[width=0.49\textwidth]{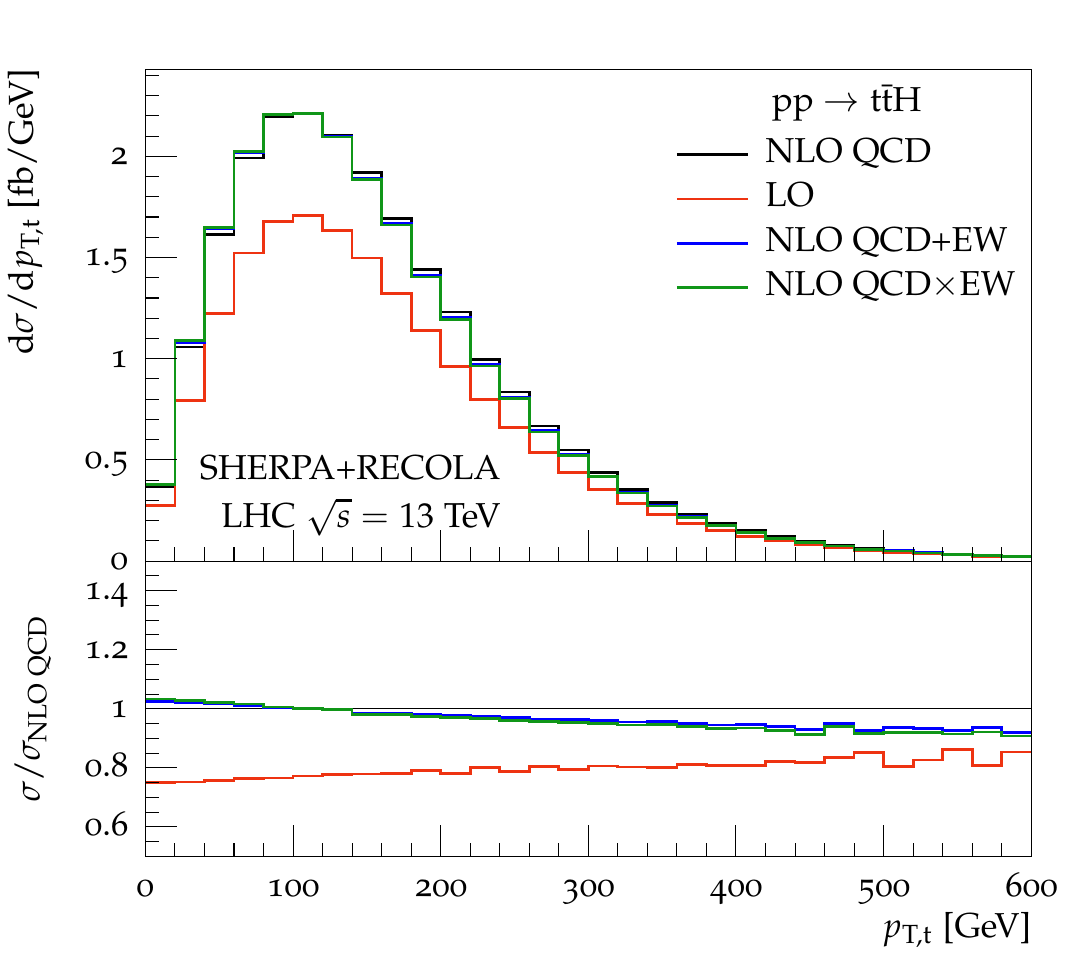}
  \includegraphics[width=0.49\textwidth]{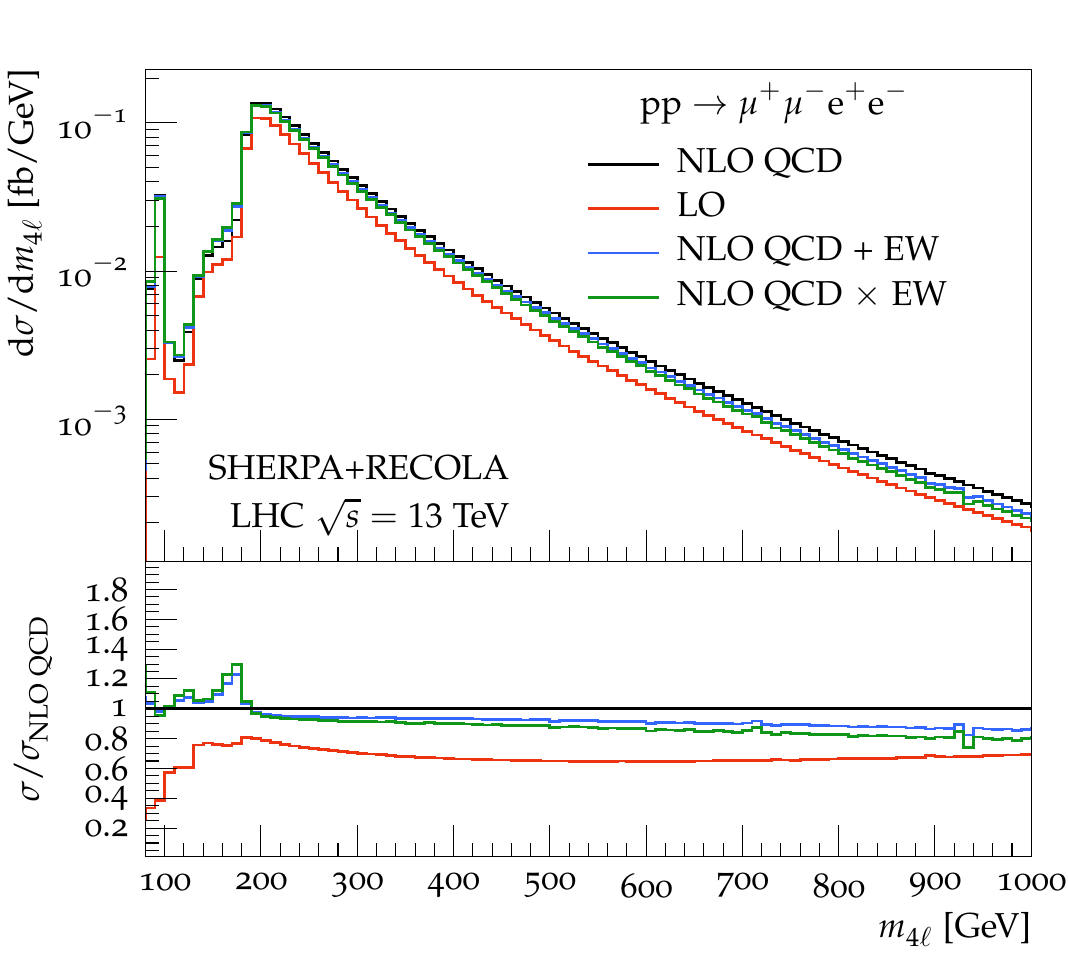}
 \end{center}
 \caption[]{Differential distributions from Ref.~\cite{Biedermann:2017yoi}: 
 transverse momentum of the top quark for ${\rm p} {\rm p} \to {\rm t} \bar {\rm t} {\rm H}$ (left) and 
 four-lepton invariant mass for ${\rm p} {\rm p} \to {\rm e}^+ {\rm e}^- \mu^+\mu^-$ (right).
 }
\label{fig:1}
\end{figure}

The first example is the on-shell production of Higgs bosons in association with two top quarks.
It has two massive coloured final states and at NLO EW, it is necessary to compute interference contributions.
The first NLO EW computation for such a process as been reported in Ref.~\cite{Frixione:2015zaa} for on-shell top quarks and in the narrow-width approximation in Ref.~\cite{Yu:2014cka}.
In Ref.~\cite{Denner:2016wet}, the NLO EW corrections to the full off-shell process have been presented.
The validation of this process has been performed against Ref.~\cite{Badger:2016bpw} where comparisons between different generators have been performed.
On the left-hand side of Fig.~\ref{fig:1}, the transverse-momentum distribution of the top quark is shown.
It displays the typical behaviour of Sudakov logarithms that grow negatively large in the tail of the distribution.
In addition, the NLO QCD corrections are also computed and combined to the EW ones with both the additive and multiplicative prescription.

\subsection{${\rm p} {\rm p} \to {\rm e}^+ {\rm e}^- \mu^+\mu^-$}

The second process is the production of two off-shell Z bosons.
There has been recently a lot of interest for the computation of EW corrections in this channel~\cite{Biedermann:2016yvs,Biedermann:2016lvg,Kallweit:2017khh}.
This process has been checked against the results of Ref.~\cite{Biedermann:2016lvg}.
It is a non-trivial process as it has four charged particles in the final state.
In particular, it possesses non-trivial kinematic edges as it can be seen on the right-hand side of Fig.~\ref{fig:1} for the four-lepton invariant mass distribution.
The lower part of the distribution displays non-trivial corrections~\cite{Biedermann:2016lvg} while for high invariant mass, the EW corrections become negatively large.
Again, the NLO EW corrections are combined with NLO QCD predictions in two ways.
It is worth noting that the results of Refs.~\cite{Biedermann:2017yoi,Kallweit:2017khh} have been obtained in a fully automatic way.
Hence, NLO QCD and EW corrections for (any) di-boson processes can now easily be obtained in the Sherpa in combination with either Recola or OpenLoops.

\subsection{${\rm p} {\rm p} \to V + {\rm jets}$}

\begin{figure}
 \begin{center}
  \includegraphics[width=0.49\textwidth]{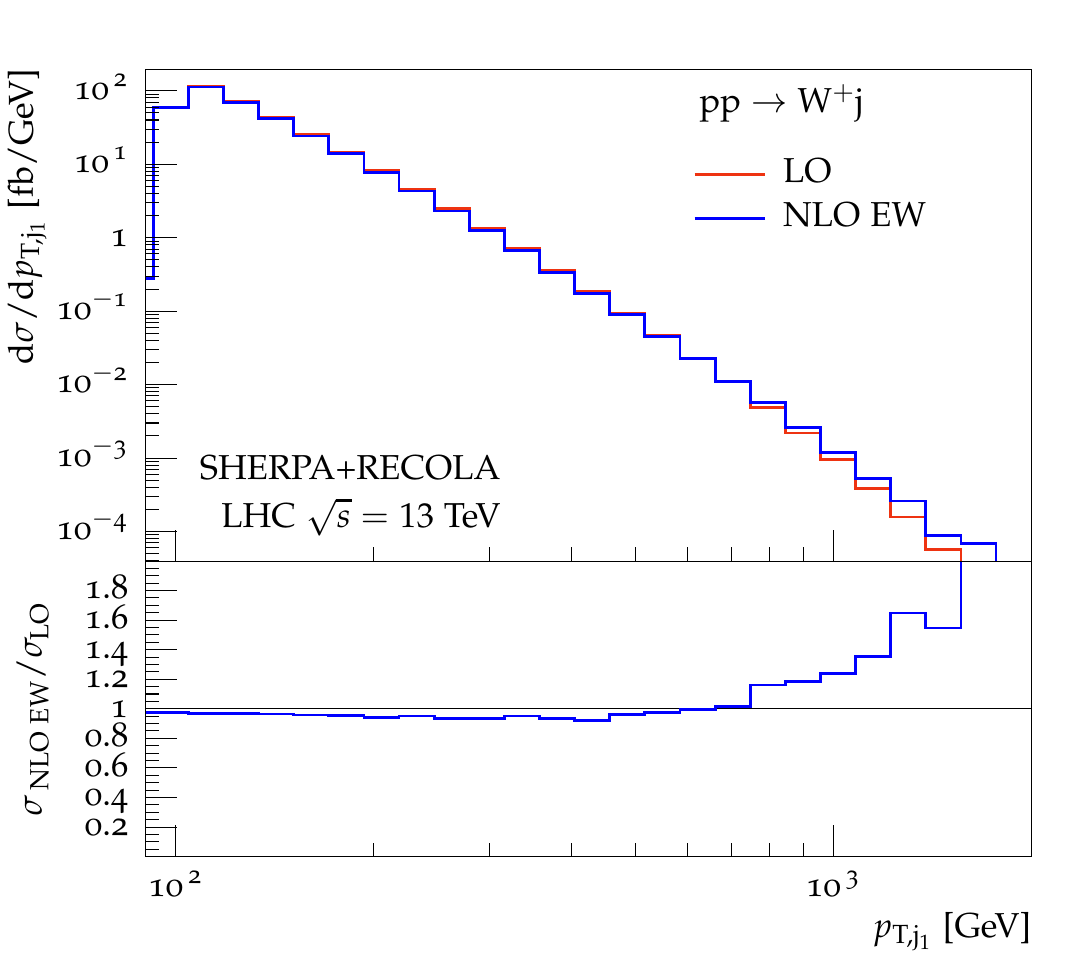}
  \includegraphics[width=0.49\textwidth]{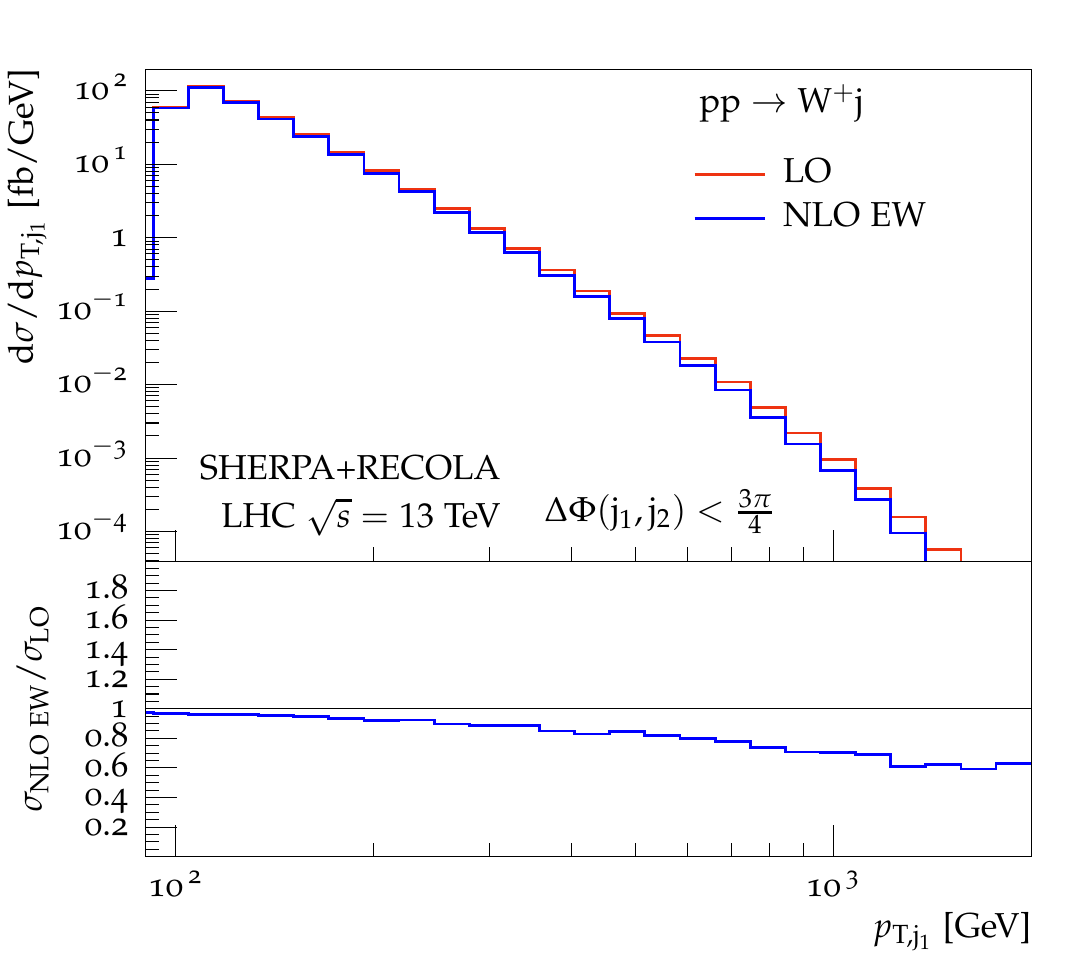}
 \end{center}
 \caption[]{Differential distributions from Ref.~\cite{Biedermann:2017yoi}:
 transverse momentum of the hardest jet without (left) and with (right) a $\Delta \Phi \left( {\rm j_1}, {\rm j_2} \right)$ cut.
 }
\label{fig:2}
\end{figure}

The last example is the production of a vector boson in association with several jets.
The state-of-the art NLO EW corrections for such processes have been presented in Refs.~\cite{Denner:2014ina,Kallweit:2014xda,Kallweit:2015dum}.
The results presented in Ref.~\cite{Biedermann:2017yoi} have been validated against the ones of Refs.~\cite{Kallweit:2014xda,Kallweit:2015dum}.
On the left-hand side of Fig.~\ref{fig:2}, the transverse momentum of the hardest jet for the process ${\rm p} {\rm p} \to {\rm W}^+ + {\rm j}$ is displayed.
This is not the usual scaling of EW corrections in the high-energy region as shown for the previous distributions on Fig.~\ref{fig:1}.
Here, the EW corrections become positively large above 1 TeV.
This behaviour is due to the appearance of back-to-back di-jet contributions in the real radiation~\cite{Denner:2014ina,Kallweit:2014xda,Kallweit:2015dum}.
Such configurations can efficiently be removed with the introduction of a $\Delta \Phi \left( {\rm j_1}, {\rm j_2} \right)$ cut.
On the right-hand-side of Fig.~\ref{fig:2}, the same distribution is shown with such a cut at a value of $4\pi/3$.
By removing these kind of topologies, one then recovers the typical behaviour of EW corrections in the high-energy limit.
Note that the plots on Fig.~\ref{fig:2} are for the process ${\rm p} {\rm p} \to {\rm W}^+ + {\rm j}$ but in the original article~\cite{Biedermann:2017yoi}, results are also shown for the production of off-shell an gauge boson in association with up to three jets.

\section{Summary}

I have presented few NLO computations done in the Sherpa+Recola framework.
It allows to compute arbitrary processes in the Standard Model with NLO QCD and EW accuracy.

\section*{Acknowledgments}

I acknowledge financial support from  BMBF under contract 05H15WWCA1.

\end{document}